\documentclass{eas}
\usepackage{graphicx}
\begin{document}

\title{Low X-ray background measurements at the Underground Canfranc Laboratory}
\runningtitle{Galan \etal: Low X-ray background \dots}

\author{J.~Gal\'an}
\address{Grupo de F\'isica Nuclear y Astropart\'iculas, University of Zaragoza, Zaragoza, Spain;
	\email{Javier.Galan.Lacarra@cern.ch}}
\secondaddress{IRFU, Centre d'\'Etudes de Saclay, CEA, Gif-sur-Yvette, France}
\author{S.~Aune}
\sameaddress{2}
\author{T.~Dafni}
\sameaddress{1}
\author{G.~Fanourakis}
\address{Institute of Nuclear Physics, NCSR Demokritos, Athens, Greece}
\author{E.~Ferrer-Ribas}
\sameaddress{2}
\author{J.A.~Garc\'ia}
\sameaddress{1}
\author{A.~Gardikiotis}
\address{University of Patras, Patras, Greece}
\author{T.~Geralis}
\sameaddress{3}
\author{I.~Giomataris}
\sameaddress{2}
\author{H.~G\'omez}
\sameaddress{1}
\author{J.G.~Garza}
\sameaddress{1}
\author{D.C.~Herrera}
\sameaddress{1}
\author{F.J.~Iguaz}
\sameaddress{2}
\author{I.G.~Irastorza}
\sameaddress{1}
\author{G.~Luz\'on}
\sameaddress{1}
\author{T.~Papaevangelou}
\sameaddress{2}
\author{A.~Rodr\'iguez}
\sameaddress{1}
\author{J.~Ruz}
\address{CERN, European Organization for Particle Physics and Nuclear Research}
\author{L.~Segu\'i}
\sameaddress{1}
\author{A.~Tom\'as}
\sameaddress{1}
\author{T.~Vafeiadis}
\sameaddress{3}
\secondaddress{Aristotle University of Thessaloniki, Thessaloniki, Greece}
\author{S.C.~Yildiz}
\address{Do\u{g}u\c{s} University, Istanbul, Turkey}
\secondaddress{Bo\u{g}azi\c{c}i University, Istanbul, Turkey}


%
\begin{abstract}

Micromegas detectors, thanks to their good spatial and temporal discrimination capabilities, are good candidates for rare event search experiments. Recent X-ray background levels achieved by these detectors in the CAST experiment have motivated further studies in the nature of the background levels measured. In particular, different shielding configurations have been tested at the Canfranc Underground Laboratory, using a microbulk type detector which was previously running at the CAST experiment. The first results underground show that this technology, which is made of low radiative materials, is able to reach background levels down to $2 \times 10^{-7}$keV$^{-1}$s$^{-1}$cm$^{-2}$ with a proper shielding. Moreover, the experimental background measurements are complemented with Geant4 simulations which allow to understand the origin of the background, and to optimize future shielding set-ups.



\end{abstract}
\maketitle

\section{The CAST experiment.}

\begin{figure}[!h]
    \centering
     \includegraphics[width = .65\textwidth, angle = 0]{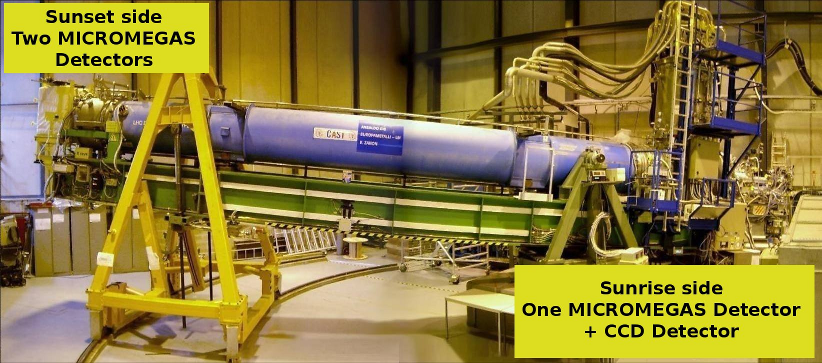}
      \caption{CAST helioscope where three micromegas detectors are taking data.}
    \label{fig:cast}
\end{figure}

Axions could be produced in the Sun via the so-called Primakoff effect. The CERN Axion Solar Telescope (CAST) experiment~(Zioutas {\em et al.\/} \cite{CAST}) uses a Large Hadron Collider (LHC) prototype dipole magnet~(Bona {\em et al.\/} \cite{magnet}) providing a magnetic field of about 9\,T which could reconvert these axions into photons when the magnetic field is transversal to the direction of axions propagation~(Sikivie~\cite{Sikivie}). The magnet aligns with the Sun twice per day (during sunset and sunrise) for about 1.5 hours each side. The magnet is composed of two independent magnetic bores whose ends are covered by X-ray detectors (see Fig.~\ref{fig:cast}). Three of the four apertures are actually covered by microbulk micromegas detectors.

The CAST research program is divided in two main phases; Phase I with vacuum inside the magnetic bore covering masses up to $m_a \leq 0.02$\,eV~(Zioutas {\em et al.\/} \cite{phase1}), and Phase II with a buffer gas inside to increase the sensitivity to higher axion masses ($m_a \leq 1$\,eV), being the magnetic bore gas density related to the enhanced axion mass~(Arik {\em et al.\/}~\cite{PhaseHe4} Aune {\em et al.\/}~\cite{PhaseHe3}). In order to achieve a competitive sensitivity on the axion-photon coupling, long data taking periods are required, specially during the second phase in which the axion parameter space ($m_a-g_{a\gamma}$) must be carefully swept in mass. Once the magnet properties are fixed, the achievable sensitivity on the axion-photon coupling is strongly dependent on the low intrinsic background of the detectors and the effect of external background radiation.

\section{CAST micromegas detectors}

The first micromegas detectors produced for CAST experiment were built by using a well established conventional fabrication process~(Giomataris~\cite{micromegas}). In these detectors, the micromesh, 3 $\mu$m thick, is made of nickel using the electroforming technique. A grid of pillars made of insulating material were mounted on the strips surface composing a precise structure of about $100\,\mu$m thick where the micromesh is placed.

\medskip

\begin{figure}[!h]
    \centering
	\begin{tabular}{ccccc}
     \includegraphics[width = .25\textwidth, angle = 0]{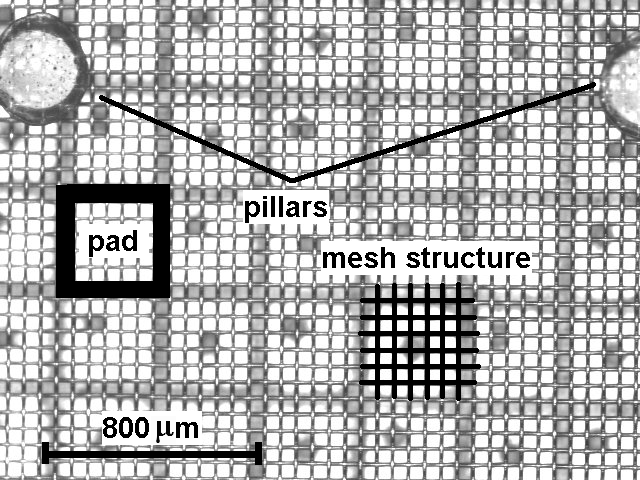}	&	&
     \includegraphics[width = .25\textwidth, angle = 0]{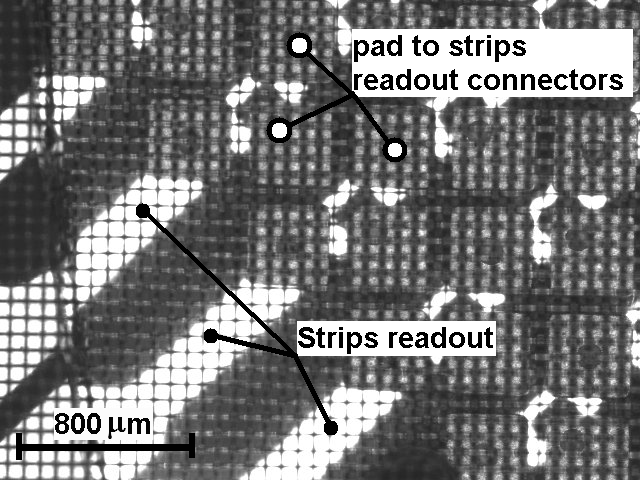}	&	&
     \includegraphics[width = .25\textwidth, angle = 0]{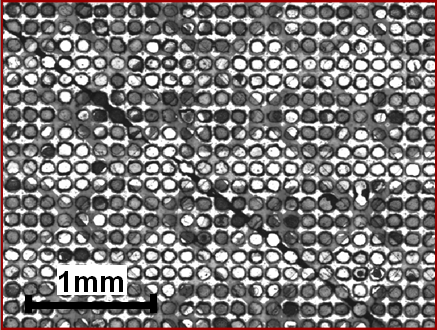}	\\
	\end{tabular}
      \caption{Left: bulk detector microscope picture focusing at the mesh level. Center: bulk detector microscope picture focusing at the strips level. Right: mesh holes in a microbulk detector. }
    \label{fi:microscope}
\end{figure}

At the micromegas conventional technology the micromesh is placed over the strips readout by mechanical means, stretching it by using some screws that exert some pressure at the mesh boundaries. The new micromegas technologies, bulk~(Giomataris~\cite{bulk}) and microbulk~(Papaevangelou~\cite{microbulk}; Iguaz~\cite{Paco_microbulk}), differ from the conventional technology in the fact that mesh and strips form a single entity, conferring robustness to the overall structure. Some microscope pictures for bulk and microbulk detectors are shown in figure~\ref{fi:microscope}.

\medskip

The micromegas detectors built for CAST allow to obtain a temporal signal coming from the mesh, and spatial information from the charge induced and integrated at the XY-strips readout. This allows to use pattern recognition techniques to distinguish between different kind of events having place in the detector.

\medskip

Low background detectors are achieved by optimizing \emph{three} main contributions. First is required \emph{low intrinsic detector radioactivity}, which is a fact in micromegas detectors due to the low amount of materials required and specially for the microbulk technology where mesh and readout strips are contained in a structure of 80$\mu$m thick made of radiopure materials, as copper and kapton\footnote{ The rest of the mechanical structure of the detector is made of Plexiglas (also radiopure) with the exception of the drift cathode which is made of stainless steel}. Radiopurity measurements have shown excellent results for the microbulk type~(Cebri\'an~\cite{paquito}). Second, a \emph{reduction of external background} by using a shielding typically made of copper, lead, Plexiglas and/or other clean materials. And third, \emph{good discrimination capabilities} to reduce the amount of additional background noise coming from cosmic rays and anything else that is not an X-ray. This last point is achieved by exploiting the temporal and spatial information of the event~(Gal\'an~\cite{myThesis}).

%
%

\section{Studies motivation, detector set-up and first background measurements.}

During the last years, the evolution of micromegas detection systems inside the CAST experiment, in terms of detectors performance and shielding upgrades, has pushed the background levels towards new lower limits. Before 2007, the first conventional micromegas detector in CAST was operating without a protecting shielding, the nominal background level reached by this set-up was around $4.5\times10^{-5}\,$keV$^{-1}$cm$^{-2}$s$^{-1}$. The improved performance achieved by the new detector technologies together with the implementation of a shielding allowed to set the background level below $1\times 10^{-5}$\,keV$^{-1}$cm$^{-2}$s$^{-1}$~(Gal\'an~\cite{Crete2009}).

\medskip

The good results obtained in the CAST experiment with micromegas detectors have pushed the efforts towards more detailed studies on the nature of the background. These studies are motivated by the possibility to reach lower limits and increase the expectation of future axion searches with a New Generation Axion Helioscope (NGAH)~(Irastorza~\cite{NGAH}).

\medskip

\begin{figure}[!h]
    \centering
	\begin{tabular}{ccc}
     \includegraphics[width = .214\textwidth]{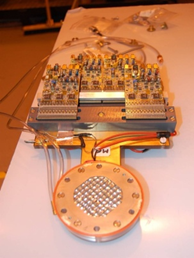}	&	&
     \includegraphics[width = .38\textwidth]{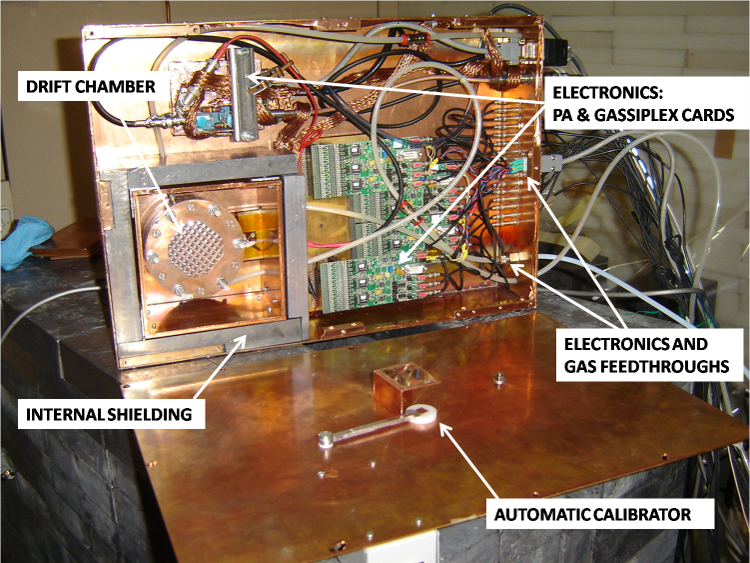}	\\
	\end{tabular}
      \caption{On the left, a stand-alone microbulk detector where the Plexiglas chamber with its stainless steel drift window, supporting structure with electronic cards and gas connections can be observed. On the right, the Faraday cage containing the detector as described in the text. }
    \label{fig:LSCset-up}
\end{figure}

A new set-up was built at the University of Zaragoza reproducing the one installed at CAST sunrise side by using an equivalent shielding configuration. A copper box was prepared to place the detector and electronic cards inside. This box was conceived with the double function of electronic noise reduction as a Faraday cage, and insulation from the environment, preserving the leak tightness of the box as good as possible by using the proper feedthroughs for electronics, gas and high voltage cables. Additionally an input entry exists at the box to flush nitrogen at the internal part of the shielding, where the chamber is placed, producing a slight overpressure and keeping a clean atmosphere around. The detector was fixed inside the cage, and the chamber was placed inside a 5\,mm thick copper frame surrounded by 2.5\,cm of lead (see Fig.~\ref{fig:LSCset-up}). The required 5\,mm copper and 2.5\,cm lead at the top and at the bottom of the detector to complete the CAST-like shielding are placed outside, once the cage is closed. An automatic calibration system, specially designed for this set-up, allows to check the detector gain and performance periodically.

The first background measurements were carried out at the surface (using the spare M13 detector from CAST) in the laboratory at the University of Zaragoza, showing levels comparable to those obtained in the detectors operating in CAST. Several environmental parameters and electronics configurations were explored in search of correlation with background levels. Only the configurations affecting the performance of the detector (such as a reduced mesh electron transmission and/or a worse energy resolution) showed a negative effect in the background level. 

\section{CAST-like set-up measurements underground}

Once the set-up and the background levels at surface were characterized, the system was moved to one of the halls at the Canfranc Underground Laboratory (LSC). The LSC is situated under 2500 m.w.e. (meter water equivalent) in the Spanish Pyrenees, where we know muons are reduced by a factor 10$^{-4}$~(Luz\'on \cite{LSCmuons}).

\begin{figure}[!h]
    \centering
     \includegraphics[width = .48\textwidth, angle = 0]{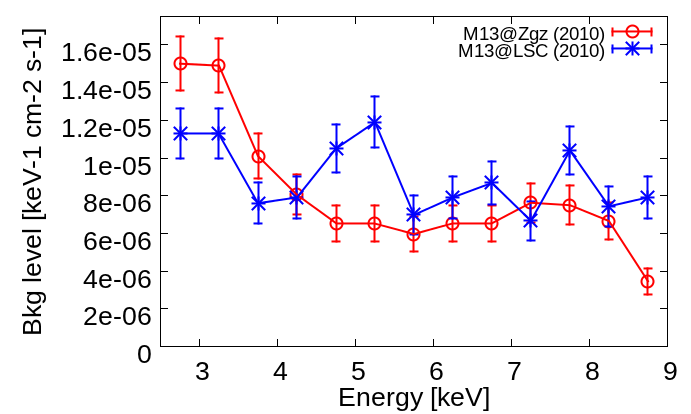}
     \includegraphics[width = .48\textwidth, angle = 0]{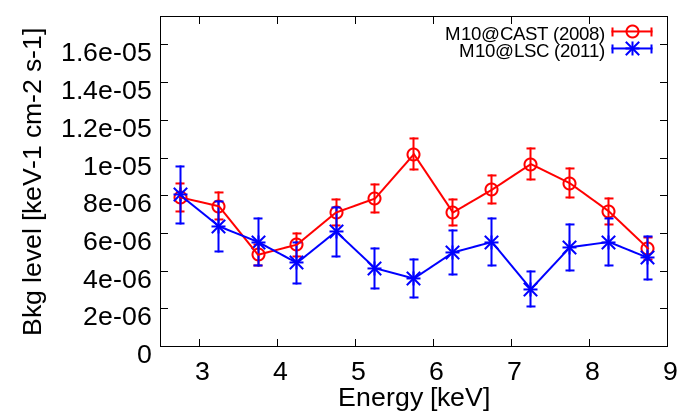}
      \caption{Left: final background spectra taken in identical conditions in the Zaragoza's test-bench (surface) and in the LSC (2500 m.w.e. deep) with the M13 detector.Right: final background spectra taken with the M10 detector in the CAST experiment (sunrise side, Autumn 2008) and in the LSC (Winter 2011) in similar shielding configurations.}
    \label{fig:ZgzvsLSC}
\end{figure}

\medskip
The effect of bringing the detector underground was observed at the trigger rate, before the offline X-ray selection is performed. The trigger rate was reduced to 0.2\,Hz, by more than a factor 5 respect to surface. The direct comparison between the background measured at surface and at the LSC, at the same conditions, showed no significative difference (see Fig.~\ref{fig:ZgzvsLSC}). Thus, proving the good discrimination capabilities of the detector which is clearly rejecting the majority of cosmic rays interacting at the surface set-up.

Afterwards, the M13 detector was replaced by the M10 detector which took data at the CAST experiment in 2008. The better performance of M10 in comparison with M13 could be appreciated by the lower background level achieved, emphasizing the importance of the detector performance on the discrimination capabilities. The background measured in the LSC could be compared directly with the data taken at the CAST experiment in running conditions (see Fig.~\ref{fig:ZgzvsLSC}). The differences found between both spectra are coming from the differences in the shielding configurations; while the underground set-up is a complete 4$\pi$-shielding, the shielding in CAST limited by the pipe connection to the magnet, which produces a weakness in the protection against external radiation. In particular, the peaks observed at the CAST background data can be attributed to the fluorescence peaks produced by external gammas in the materials surrounding the detector, mainly stainless steel (Cr, Fe, Ni at 5.4, 6.4 and 7 keV) and copper. These peaks are significantly reduced at the LSC by a more efficient gamma suppression at the inner region.


\section{Shielding upgrade underground.}

\begin{figure}[!h]
    \centering
     \includegraphics[width = .8\textwidth, angle = 0]{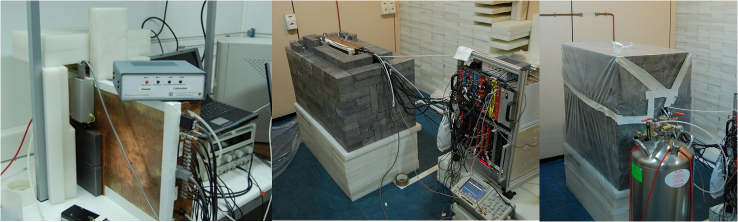}
      \caption{Three different configurations for the external lead shielding, from left to right: CAST-like 2.5 cm lead; 20 cm extra lead layer without closing the top of the lead castle and completing the external shielding. Continuous nitrogen flow is present in all three configurations.}
    \label{fig:shielding_upgrades}
\end{figure}

After the characterization of the CAST-like shielding, the shielding was increased by adding an additional wall of lead of 20 cm in all directions. By growing the shielding underground we are avoiding the negative effect that this kind of shielding could have at surface, due to the effect of secondaries which could be produced by interactions in the shielding itself. The upgrade was carried out in \emph{two} steps, first covering bottom and lateral walls, and afterwards completing the shielding with the top wall (see Fig.~\ref{fig:shielding_upgrades}).

\begin{figure}[!h]
    \centering
     \includegraphics[width = .8\textwidth, angle = 0]{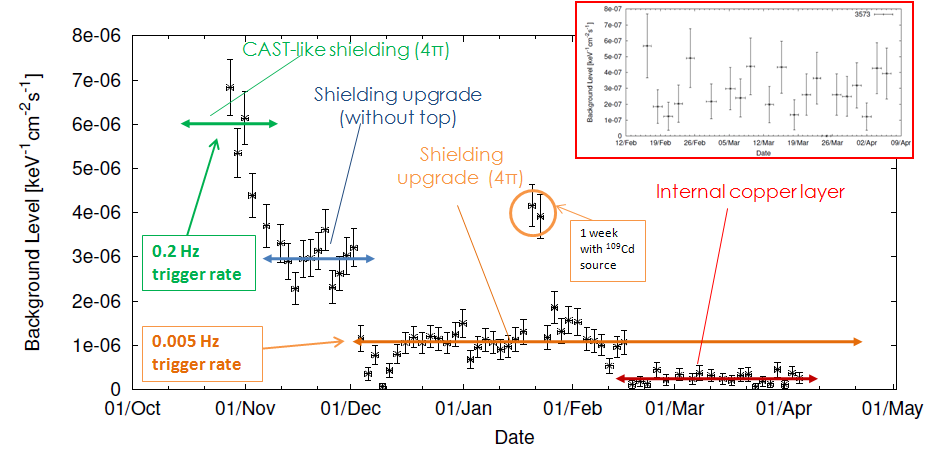}
      \caption{Evolution of the final background level of the M10 detector in the LSC along different set-up upgrades. The ultra-low background period is zoomed.}
    \label{fig:bkgprogression}
\end{figure}

The two-steps shielding upgrade affected clearly at the background level achieved by the detector (see Fig.~\ref{fig:bkgprogression}). Once the shielding was completed a factor 6 reduction was obtained, and a level of about $1\times10^{-6}$keV$^{-1}$cm$^{-2}$s$^{-1}$ was reached. More impressive was the reduction observed at the trigger rate by a factor $\simeq 40$, indicating that the majority of the interactions in the chamber have an external origin.

There were reasons to suspect that the background level achieved at this first stage was being limited by an internal contamination, due to the clear presence of a peak at the 5-6 keV region that could be produced by the $^{55}$Fe placed inside the cage, and that now was observable thanks to the lower background levels reached. In order to screen this contamination from the detector field of view, an additional copper layer was placed on top of the detector window by using different configurations~(Tom\'as~\cite{Alfredo_microbulk}). The final result was the obtention of a more restrictive limit on the intrinsic background of the detector at a level of $1.5\pm0.6$\,keV$^{-1}$cm$^{-2}$s$^{-1}$ at the 2-7 keV energy range, value which was obtained after a total exposure time of 994 hours.

\section{Summary and conclusions}

\begin{figure}[h!]
    \centering
     \includegraphics[width = .43\textwidth, angle = 0]{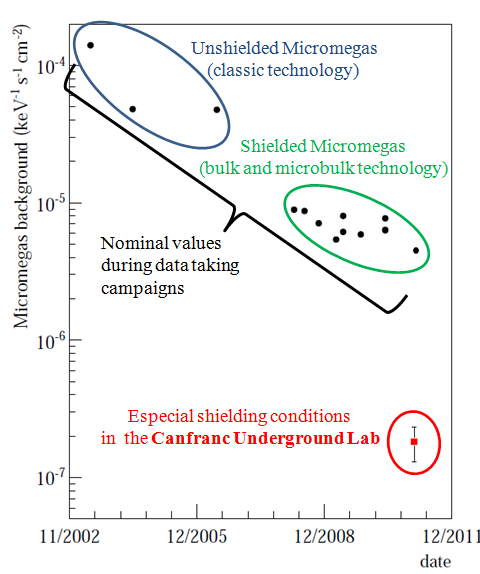}
      \caption{Historical background levels for CAST micromegas detectors, together with the lower value obtained in this work. }
    \label{fig:history}
\end{figure}

We have presented the importance of low X-ray background detectors for rare event searches. The advances towards lower background detectors affects directly to the experiment involved increasing its sensitivity. We have shown that micromegas microbulk technology, with the proper shielding and conditions, can achieve even lower background levels than those that are actually providing data in the CAST experiment, levels which have also shown considerable improvement during the different phases of the experiment (see Fig.~\ref{fig:history}). 

Moreover, a complete methodology and set-up to evaluate future low intrinsic radiation detectors has been well established. Future work is in the course to produce new detectors with even lower radiative levels. This experimental search together with the understanding of the background nature and optimization of the shielding thanks to detailed Geant4 simulations could provide the key to obtain even lower limits.

\section*{Acknowledgments}
We acknowledge support from the European Commission under the European Research Council T-REX Starting Grant ref. ERC-2009-StG-240054 of the IDEAS
program of the 7th EU Framework Program, as well as, from the Spanish Ministry of Science and Innovation (MICINN) under contract ref. FPA2008-03456, and the CPAN project ref. CSD2007-00042 from the Consolider-Ingenio 2010 program of the MICINN. Part of these grants are funded by the European Regional Development Fund (ERDF/FEDER).



\begin{thebibliography}{}

\bibitem[2009]{PhaseHe4}
Arik E {\em et al\/}. [CAST Collaboration] 2009, JCAP, {\bf 0902} 008 [arXiv:0810.4482 [hep-ex]]

\bibitem[2011]{PhaseHe3}
Aune S {\em et al\/}. [CAST Collaboration] 2011, [arXiv:1106.3919]. To be published in Phys. Rev. Lett

\bibitem[1994]{magnet}
Bona M {\em et al\/}. 1994, Performance of the first CERN-INFN 10\,m long superconducting dipole prototype for the LHC, CERN-LHC-Note-276

\bibitem[2011]{paquito}
Cebri\'an S {\em et al.} 2011, Astropart. Phys., {\bf 34}, 354 [arXiv:1005.2022v1]

\bibitem[2010]{Crete2009}
Gal\'an J {\em et al.} 2010, JINST \textbf{5}, P01009

\bibitem[2011]{myThesis}
Gal\'an J, 2011, {\it PhD Thesis}, Univ. de Zaragoza [arXiv:1102.1406]

\bibitem[1996]{micromegas}
Giomataris Y, Rebougeard P, Robert JP, Charpak G, 1996 Nucl.\ Instr.\ and Meth.\ A, {\bf376}, 29

\bibitem[2006]{bulk}
Giomataris Y {\em et al.} 2006, Nucl.\ Instr.\ and Meth.\ A, {\bf 560} 405

\bibitem[2011]{Paco_microbulk}
Iguaz FJ {\em et al.} 2011, TIPP Conference 2011, to be published in Physics Procedia

\bibitem[2011]{NGAH}
Irastorza I G {\em et al.} 2011, JCAP \textbf{06}, 013

\bibitem[2010]{microbulk}
Papaevangelou T {\em et al.} 2010, JINST \textbf{5} P02001

\bibitem[1983]{Sikivie}
Sikivie P. 1983 Phys. Rev. Lett. \textbf{51}, 1415

\bibitem[2011]{Alfredo_microbulk}
Tom\'as A {\em et al.} 2011, TIPP Conference 2011, to be published in Physics Procedia

\bibitem[2006]{LSCmuons}
Luz\'on G {\em et al.} 2006, in {\it Proceedings of the International Conference in the Identification of Dark Matter}, p. 514

\bibitem[1999]{CAST}
Zioutas K {\em et al.} 1999, Nucl. Instrum. and Meth. A, {\bf 425}, 27.

\bibitem[2007]{phase1}
Zioutas K {\em et al.} [CAST Collaboration] 2005, Phys. Rev. Lett., {\bf 94}, 121301 [arXiv:hep-ex/0411033]
\end{thebibliography}
\end{document}